# Spherical Manifold in Quantum Evolution

Aalok Pandya* and Ashok K. Nagawat

Department of Physics, University of Rajasthan, Jaipur 302004, India.

**Abstract**

In this paper, the projective geometry is used to describe the features of spherical manifold and discreteness in quantum evolution. As a system evolves in time the state vector changes and it traces out a curve in the Hilbert Space. Geometrically, the evolution is represented as a closed curve in the Projective Hilbert space. In recent times many attempts have been made to describe 'length', 'distance' and 'geometric phases' and also 'parallel transport' and symplectic geometry in various ways in the Projective Hilbert space, during quantum evolution. It is shown in this paper that for the quantum evolution in ray space, spherical manifold and features of discreteness can be described geometrically.

___________________

*E-mail: aalok_cosmos@yahoo.com

# 1. Introduction

There is a growing feeling that quantum mechanics is fundamentally a geometric theory [1]. The interest of scientific community in the geometric aspect of quantum evolution was invoked with the advent of Berry's famous paper on geometric phases [2]. The geometry of the quantum evolution is illustrated by means of connections and symplectic structures associated with Projective Hilbert space $\mathscr{P}$[1]. As a system evolves in time the state vector changes and it traces out a curve in the Hilbert space $\mathscr{H}$. Geometrically the evolution is represented as a closed curve in the Projective Hilbert space [1, 3-9]. In recent times many attempts have been made to describe length, metric, and geometric phases in the projective Hilbert space, during quantum evolution [1-11]. Also, manifold, parallel transport and holonomy have been described in the studies of quantum evolution [1, 10-14]. In the same year in which Berry discovered geometric phases, Simon [12] explained that this geometric phase could be viewed as a consequence of parallel transport of vectors in a curved space appropriated to the quantum system. It appears naturally as due to holonomy of a line bundle over the parameter space [12]. The problem of describing quantum mechanics on a manifold had been attempted earlier too [15-24]. In this paper spherical geometry has been described in quantum evolution. This paper describes analytic geometric work in the context of presently ongoing work on quantum evolution, in particular [1-14]. In addition, our work in this paper is different from the earlier work in the sense that it picturizes the projective geometry in the ray space. But here, only the geometric aspect has been considered and kinematics has been excluded. Illustrated mappings (figures 2, 3, 4) are used for the geometric treatment. A set of rays is regarded as a physical object of discussion. However, the authors do not claim to produce any new mathematics but the whole exercise is to attribute a mathematical and geometric explanation to an idea of physical relevance.

## 2. Wave function and manifold characteristics

As the system evolves in time the state vector changes and it traces out a curve in the Hilbert space $\mathcal{H}$. For cyclic evolution this is an open curve **C**. However, its projection is a closed curve $\hat{C} = \Pi(\mathbf{C})$. Therefore, geometrically the evolution is represented as a curve in the projective Hilbert Space $\mathcal{P}$. Where the set of rays of the Hilbert space forms a projective Hilbert space $\mathcal{P}$ via some projective map [4-9]. This could be analyzed in yet another way with the help of Fubini Study Metric [4-9, 27]. Fubini Study metric is given by

$$ds^2 = 4(1 - |\langle \Psi_1 | \Psi_2 \rangle|^2) \tag{1}$$

or $$ds^2 = 4(1 - |\langle \Psi(t) | \Psi(t+dt) \rangle|^2) \tag{2}$$

Here, $s$ represents a geodesic on the sphere $P_1(C)$, the projective space of the subspace spanned by $|\Psi(t_1)\rangle$ and $|\Psi(t_2)\rangle$, which is an arc of a great circle [1]. It can be shown in many ways that $s$ represents a circle on $CP_1$. Here, the wave function $\Psi$ obeys Schrodinger picture. Now, we take a wave function in a specific case;

$\Psi_{1,1} = -\left(\dfrac{3}{8\pi}\right)\dfrac{x+iy}{r}$, so that the corresponding conjugate function $\Psi_{1,-1}$ can be given as $\Psi_{1,-1} = \left(\dfrac{3}{8\pi}\right)\dfrac{x-iy}{r}$. (3)

Or alternatively, we can also write the function by permutation of $x, y,$ and $z$, at any particular instant, such that: $\Psi^*\Psi = k^2$ (say). (4)

Where, $k$ is any number. Equation (3) with (4) implies

$x^2 + y^2 = \left(\dfrac{9}{64\pi^2}\right)k^2 r^2$; or, if we equate $\left(\dfrac{3k}{8\pi}\right)^2 r^2 = R^2$, we can write this equation as: $x^2 + y^2 = R^2$. (5)

This is an equation of a circle and circle is a manifold $S^1$. However, in general we could take wave function in such a form that it could imply equation of an ellipse also. *But, we stick to the above choice for the sake of mathematical simplicity, and without any loss of generality, an ellipse can always be conformally mapped to a circle and vice-versa.* We will have further analytical explanation for this in the last section of this paper. We can see that there exists a sub-bundle $\mathfrak{S}$ such that,

$\mathfrak{S} = \{\,|\Psi\rangle \in \mathcal{H}.\ \langle\Psi|\Psi\rangle = 1\,\}$ of $\mathcal{H}^*$.

And $\mathfrak{S}$ is the Hopf bundle over $\mathcal{P}$. As naturally $\mathcal{H}$ has finite dimensions N then $\mathcal{P}$ has dimensions N-1 [Ref 3]. In the next section we shall consider mapping by transforming co-ordinates as $(x, y) \to (Z, \bar{Z})$ i.e. in this section we considered the wave function at time, t = 0 and now in the next section we want to study it with time evolution.

## 3. Lobachevskian spherical metric

We consider a Euclidean plane and introduce on it a "complex co-ordinate" Z= x+iy; and $\bar{Z}$ = x-iy. Here we shall not concentrate on the geometric meaning of the new co-ordinates; but consider the mapping $(x,y) \to (Z,\bar{Z})$ as a formal co-ordinate transformation. The Jacobi matrix of this transformation is;

$$J = \begin{pmatrix} 1 & i \\ 1 & -i \end{pmatrix}; \text{ Hence the Jacobi J } = -2i \neq 0. \tag{6}$$

Thus the transformation is regular.

Since $dZ = dx + i\, dy$ and

$$d\bar{Z} = dx - idy, \tag{7}$$

The Euclidean metric in these new co-ordinates takes the form;

$$ds^2 = dx^2 + dy^2 = (dx + i\, dy)(dx - i\, dy) = dZ\, d\bar{Z} \tag{8}$$

Hence, the metric of a sphere is;

$$ds^2 = \frac{4(dx^2 + dy^2)}{(1 + x^2 + y^2)^2}$$
$$= \frac{4 dZ d\bar{Z}}{(1 + |Z|^2)^2} \tag{9}$$

Where, $|Z|^2 = Z.\bar{Z} = x^2 + y^2 = r^2$. (10)

In nutshell, a circle is under consideration, which is embedded on a point. And this embedding describes a spherical surface and hence a spherical metric. Thus a propagating wave packet in the ray space, sweeps a spherical metric all the way. But of course, the spherical metric described here is without the North and South

Pole. $\mathscr{H}^* = \mathscr{H} - \{0\}$ is a principle fiber bundle over $\mathscr{P}$ with structure group C* (the group of non zero complex numbers), and the disjoint union of rays in $\mathscr{H}$ is the natural line bundle over $\mathscr{P}$ whose fiber above any p∈ $\mathscr{P}$ is p itself. And a sub-bundle $\mathfrak{J}$ is such that;

$$\mathfrak{J} = \{ |\Psi\rangle \in \mathscr{H} \langle\Psi|\Psi\rangle = 1 \} \text{ of } \mathscr{H}^*.$$

$\mathfrak{J}$ is the Hopf bundle over $\mathscr{P}$. And obviously singular points described here, lie on the axis of propagation along the poynting vector. These points are none other than the nodes of the propagating wave. Also we like to mention here that in general, particle characteristics of the radiation field are identified with momentum and the polarization vectors [25], but we note here that localization of wave function within a spherical metric, can also be an alternative way of explaining the particle attributes. *For the entire geometrical treatment, which we have undertaken here, the most necessary requirement is that of the smoothness and continuity for all kind of embedding and mappings, otherwise we won't be able to define a manifold. And in the context of our description, this condition is readily fulfilled. As for the wave function we know that it's always continuous in nature.* In the next section we arrive at similar remarks but in a much more rigorous way.

## 4. Spatial analysis and manifold description

Let us consider propagation of a set of rays, for example it can be on an electro magnetic wave, where in electric and magnetic vectors are perpendicular to each other and the poynting vector i.e. direction of propagation is perpendicular to both of these. The figures (1 & 2) of ray propagation and its cross sectional view enables us to have a rigorous description by analytical methods. With the points under consideration (figure 2) we establish mappings as illustrated. Here locally homeomorphism is observed within a plane or between the perpendicular planes. If a metric space **M** is an n- dimensional manifold we can find in **M,** a system of open sets $\{U_i\}$ numbered by finitely (or infinitely) many indices *i* and a system of homeomorphisms $\Phi_i : U_i \to V_i \subset R^n$ of the sets $U_i$ on the domain $V_i$. The system $\{U_i\}$ must cover the space **M,** i.e. $\mathbf{M} = \cup U_i$, and the domain $V_i$ may, in general, intersect one another. Suppose a Cartesian co-ordinate system $(x^1,..........,x^n)$ is valid in a Euclidean space $\mathbf{R}^n$, **t**hen for any point $P \in U_i$, the Cartesian co-ordinates of the point $\Phi_i(p) \in V_i$ can be considered as a numerical parameterization of P. The homeomorphism $\Phi_i$ is therefore a co-ordinate homeomorphisms and the Cartesian co-ordinates $(x^1,............, x^n)$ termed as local co-ordinates of the point $P \in U_i$ and denoted by; $x^k = x^k(p)^*$, with k =1, 2 ,............n. A system of functions, $x^k = x^k(p)^*$ given on an open set $U_i$ is a local co-ordinate system, and the open set $U_i$ together with a local co-ordinate system defined on it, is the chart of manifold **M**. Thus the chart is pair $(U_i, \Phi_i)$, and we denote it, for brevity only by the first symbol, $U_i$. A set of charts covering the entire manifold **M** is the atlas. It is convenient to number local co-ordinates of a point P ∈ **M** by an additional index characterizing the chart $U_i : x_i^k = x_i^k(p)$. Since the point P can belong simultaneously to several charts, it acquires several sets of local co-ordinates.

Example: We consider a circle $S^1 \subset R^2$ defined by the equation, $x^2 + y^2 = 1$.

Let us cover $S^1$ with an atlas consisting of four charts (figure 3).

$$U_1 = \{(x,y) \in S^1 : y > 0\},\ U_2 = \{(x,y) \in S^1 : y < 0\},$$
$$U_3 = \{(x,y) \in S^1 : x > 0\},\ U_4 = \{(x,y) \in S^1 : x < 0\}.$$

The corresponding domains $V_1, V_2, V_3$ and $V_4$ on the real axis $\mathbf{R}^1$ coincide and are equal to the open interval (-1,1). Homeomorphisms $\Phi_1$ and $\Phi_2$ are constructed as projections of the circle onto the x- axis: $\Phi_1(x, y) = \Phi_2(x, y) = x$, and homeomorphisms $\Phi_3$ and $\Phi_4$ as projections onto the y - axis: $\Phi_3(x, y) = \Phi_4(x, y) = y$. In order to prove that the mappings: $\Phi_k$, k = 1,....., 4 are homeomorphisms, it is sufficient to write explicitly the inverse mappings,

$$\Phi_1^{-1}(x) = \left[x, \sqrt{(1-x^2)}\right] \in S^1,\ \Phi_2^{-1}(x) = \left[x, -\sqrt{(1-x^2)}\right] \in S^1$$
$$\Phi_3^{-1}(y) = \left[\sqrt{(1-y^2)}, y\right] \in S^1,\ \Phi_4^{-1}(y) = \left[-\sqrt{(1-y^2)}, y\right] \in S^1,$$

and demonstrate that they are continuous. Then we obtain on the circle for local co-ordinate systems, each consisting only of one co-ordinate:

$x_1 = \Phi_1(x, y) = x,$     $x_2 = \Phi_2(x, y) = x,$     $x_3 = \Phi_3(x, y) = y,$

$x_4 = \Phi_4 = (x, y) = y$.

Certain points are provided with local co-ordinate systems. For instance, for point P of the intersection $U_1 \cap U_2$ the co-ordinates $x_1(p)$ and $x_3(p)$ are valid (figure 3). There are other ways of introducing an atlas on a circle. We consider polar (r, θ) on a plane. The equation of a circle in these co-ordinates is very simple: r = 1. Strictly speaking polar co-ordinates on a plane are not a co-ordinate system. We introduce therefore two charts on a circle $S^1$;

Namely $U_1 = \{(x,y) \in S^1 : x = -1\}$ and $U_2 = \{(x,y) \in S^1 : x = 1\}$ (figure 4).

Let $\Phi_1(p) = \Phi_1(x,y)$ be the value of $\Phi$ in the interval (-π, π) and $\Phi_2(p) = \Phi_2(x,y)$ be the value of $\Phi$ in the interval (0, 2π), i.e. $V_1 = (-\pi, \pi)$, $V_2 = (0, 2\pi)$. Obviously, the local co-ordinates $\Phi_1 = \Phi_1(p)$ and $\Phi_2 = \Phi_2(p)$ coincide for upper semicircle and do not coincide for the points of the lower semicircle, i.e., Y>0, $\Phi_1(x,y) = \Phi_2(x,y) - 2\pi$ for Y<0 (figure 4).

In nutshell, the mapping is one-one and onto, so it is bijection. The mapping is continuous for which inverse mapping also exists, and so it is homeomorphism. And we know that n- dimensional sphere is n-dimensional manifold. Therefore we have achieved the attributes of spherical manifold. Similar treatment can be done in the three dimensions also. Thus, we considered a classical picture of transverse wave in ray –space and we described a smooth spherical manifold in that. After describing a smooth spherical surface, the following discontinuity at the pole, reminds us of *discreteness*, which is an essential feature of quantum mechanics.

This section of the paper obviously contains all the standard Mathematics, but our innovation is to connect it with the underlying physical explanation.

## 5. Discussion and summary

The very purpose of this work is to identify more and more geometry of quantum evolution in continuation of the ongoing work on geometry of quantum evolution. In recent times many attempts have been made to describe 'length', 'distance' and 'Geometric Phases' etc. in the Projective Hilbert space, during quantum evolution. In this paper we have described spherical manifold and discreteness in quantum evolution geometrically. Summarily, having inspired by the quantization of classical systems where the dynamical variables of classical mechanics are elevated to quantum mechanical operators and then the quantization follows, we take here classical picture of transverse wave and then by defining spherical manifold therein, we get the impression of discreteness. This is an attempt towards quantization by purely geometric explanation.

## Acknowledgement

Authors are grateful to Prof. Ghanshyam Date (IMSc Chennai) and Prof. Joseph Samuel (Raman Research Institute, Bangalore) for their valuable suggestions. Also, authors wish to thank Prof. Sardar Singh and Prof. N. K. Sharma, for their suggestions and encouragement.

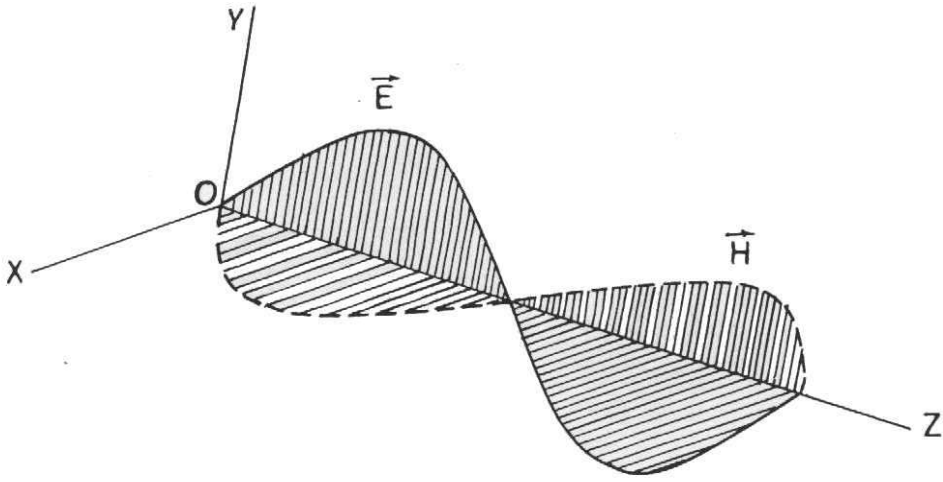

**Fig. 1 A Set of Rays in a Wave**

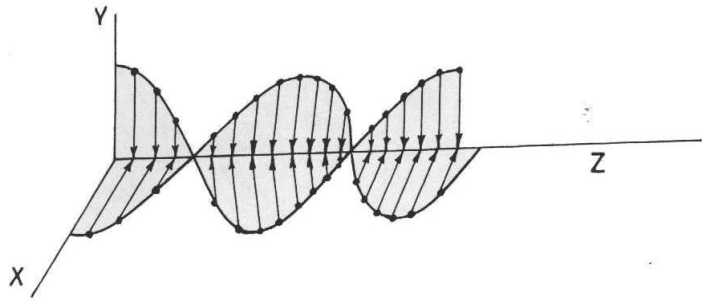

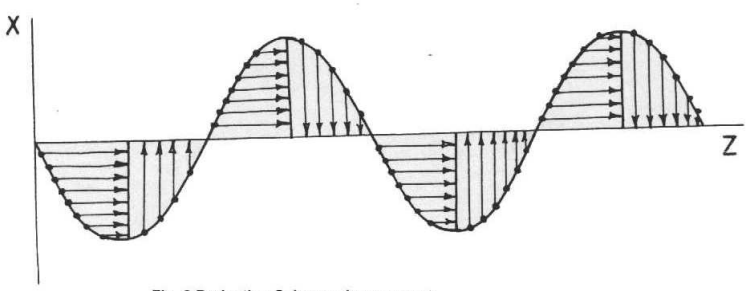

**Fig. 2 Projection Scheme along a wave**

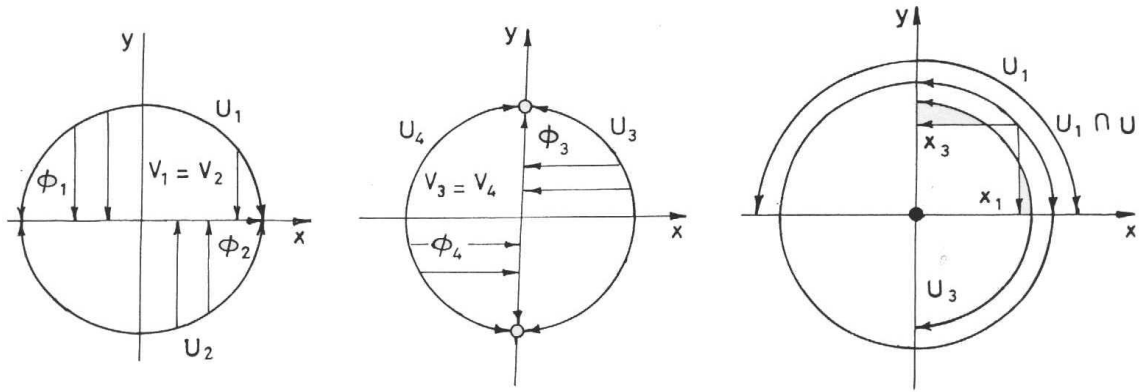

Fig. 3 Covering $S_1$ with an Atlas consisting of Charts $U_1, U_2, U_3$, and $U_4$

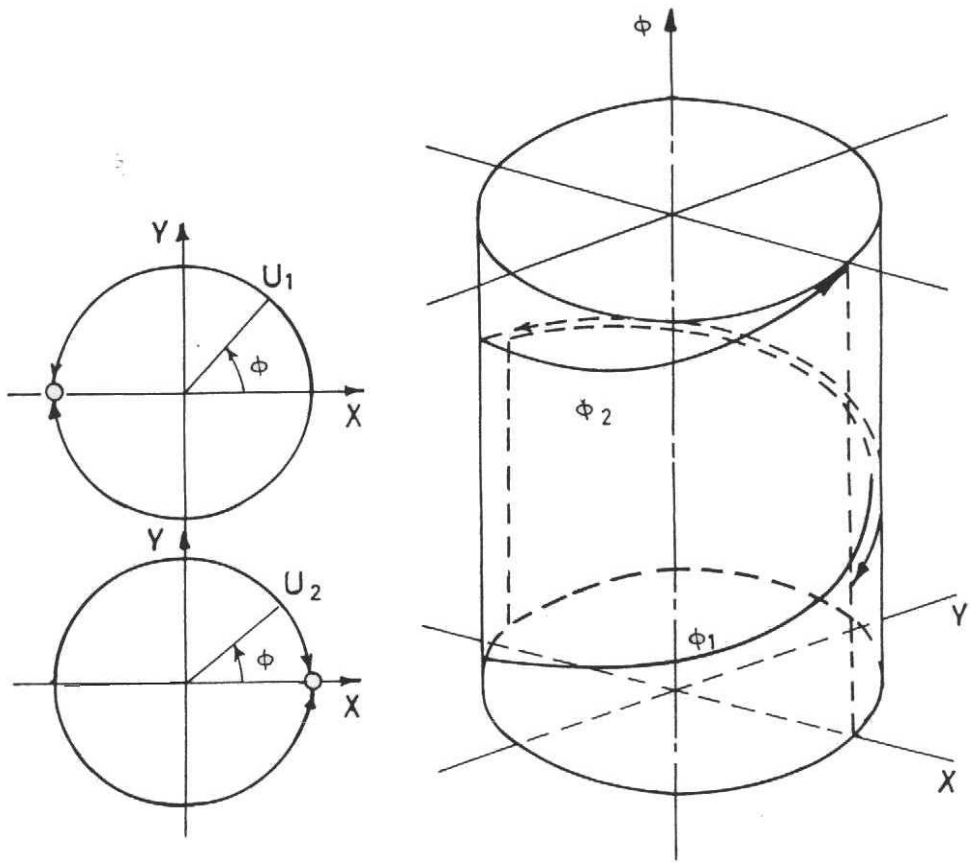

Fig. 4 : Charts $U_1$, $U_2$ and mappings $\phi_1$, $\phi_2$

# Figure Captions

FIG.1 - A set of rays in a wave.

FIG. 2 - Projection scheme along a wave.

FIG. 3 - Covering $S^1$ with an atlas consisting of charts $U_1, U_2, U_3$ and $U_4$

FIG. 4 - Charts $U_1, U_2$ and mappings $\Phi_1$ and $\Phi_2$